# Enhanced Digital Twin for Human-Centric and Integrated Lighting Asset Management in Public Libraries: From Corrective to Predictive Maintenance – A Demonstration Design

Jing (Janet) Lin[1,2*], Jingchun Shen[3,4]

1. Division of Operation and Maintenance, Luleå University of Technology, 97187, Luleå, Sweden
2. Division of Product Realization, Mälardalen University, 63220, Eskilstuna, Sweden
3. Division of Construction Technology, Dalarna University, 79188, Falun, Sweden
4. Sustainable Energy Research Centre, Dalarna University, 79188, Falun, Sweden

**Abstract:** Lighting asset management in public libraries has traditionally been reactive, focusing on corrective maintenance—addressing issues only when failures occur. Although standards now encourage preventive measures, such as incorporating a "maintenance factor," the broader goal of human-centric, sustainable lighting systems requires a shift toward predictive maintenance strategies. This study introduces an enhanced digital twin model designed for the proactive management of lighting assets in public libraries. By integrating descriptive, diagnostic, predictive, and prescriptive analytics, the model enables a comprehensive, multi-level view of asset health. The proposed framework supports both preventive and predictive maintenance strategies, allowing for early detection of issues and the timely resolution of potential failures. In addition to the specific application for lighting systems, the design is adaptable for other building assets, providing a scalable solution for integrated asset management in various public spaces.

## 1. Introduction

Traditional asset management typically prioritizes factors such as functionality, cost, and lifespan. In contrast, human-centric asset management emphasizes social sustainability and focuses on user well-being. For lighting assets, this approach considers not only the functional aspects but also aims to "promote effective lighting and provide users with the right light at the right time to support various activities" [1] [2] [3] [4].

Current lighting asset management practices in public libraries primarily rely on corrective maintenance—fixing issues only when they arise. While recent standards emphasize the importance of preventive strategies [5], a fully human-centric, sustainable approach necessitates moving beyond prevention to prediction. A predictive maintenance strategy is essential for achieving optimal asset performance and ensuring user well-being in public environments.

This study develops an enhanced digital twin model for lighting asset management that transcends the traditional focus on individual components or design/plan purpose [6, 7, 8, 9], instead emphasizing human-centric performance across different library areas. The model supports a transition from corrective to predictive maintenance, leveraging real-time data to forecast future asset performance.

Furthermore, this digital twin design is flexible and scalable, with potential applications not only in public libraries but also for various other building assets, supporting broader efforts toward sustainable infrastructure management.

---

[*] Corresponding Author: Janet (Jing) Lin

This paper is organized to showcase the design and application of an enhanced digital twin model for lighting asset management in public libraries. After the **Introduction**, the **Design Framework** section details the transition through descriptive, diagnostic, predictive, and prescriptive analytics, highlighting each phase's role in improving asset management. In the **UI Design and Functionality** section, the hierarchical interface is explained, showing how system health can be monitored at different levels, from building-wide to individual components. The paper concludes by demonstrating the scalability of the model, suggesting its broader application to various building assets.

## 2. Design Framework: From Descriptive to Prescriptive Analytics

The design framework for the digital twin model transitions through several key analytics phases: **Descriptive**, **Diagnostic**, **Predictive**, and **Prescriptive** analytics. Each phase plays a crucial role in the overall system evolution, from understanding past events to recommending future actions for optimal lighting asset management.

- **Descriptive Analytics (Monitoring)**: This phase focuses on discovering and describing what happened in the past, offering insights into the system's historical performance. It helps explain why certain events or faults occurred, providing a foundation for further investigation.

- **Diagnostic Analytics**: Diagnostic analytics delves deeper into the reasons behind faults or failures, answering the "why" and "where" questions. According to EN13306, diagnosis involves fault detection, identification, and localization, pinpointing the exact source of issues in the lighting system.

- **Predictive Analytics**: The predictive phase estimates what will happen in the future by forecasting potential system failures or performance issues. In the context of lighting asset management, predictive analytics aims to answer, "What will happen in the future?" This allows maintenance teams to anticipate problems before they occur.

- **Prescriptive Analytics**: This final phase moves beyond prediction to determine the best course of action, answering the question, "What needs to be done next?" In lighting asset management, prescriptive analytics guides decision-making for maintenance actions, offering strategies to optimize performance and reduce downtime

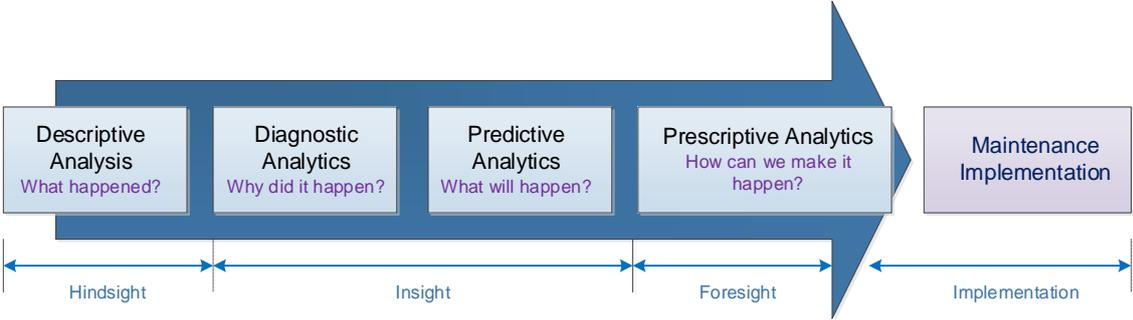

Fig.1 The framework of design

## 3. UI Design and function instruction

The User Interface (UI) is structured to provide a hierarchical view of system health and performance, structured as follows:

- **System Level (Building):** This top-level view presents an overview of the entire building's systems and their status, allowing users to monitor overall performance.

- **Subsystem Level I (e.g., Lighting System, HVAC):** At this level, users can dive into specific systems within the building, such as lighting, HVAC, or fire protection, to assess their health and functionality.

- **Subsystem Level II (e.g., Lighting Assets on specific Floor):** This view provides a more detailed look at subsystems, such as lighting assets located on specific floors, allowing users to identify localized issues or monitor performance in targeted areas.

- **Component Level (e.g., Adult Reading Area, Book Return):** The most granular view, offering insights into individual components within subsystems, such as specific areas or assets in the library. This helps in pinpointing performance issues and optimizing asset management.

- **Diagnosis, Prognosis, and Maintenance Decision Support:** The UI also incorporates diagnostic tools to identify issues, predictive analytics to forecast potential problems, and prescriptive analytics to guide maintenance decisions, ensuring seamless and efficient asset management.

## 3.1 System level (building) Monitoring

Figure 2 illustrates health status at the system (building) level.

Figure 2.1 highlights contextual information about the building, including its geographical location, current weather conditions, and time. These factors serve as examples of variables that can influence the description and diagnosis of the lighting asset health status. Additional contextual elements relevant to lighting asset management in public libraries can be found in reference [1]. These contextual factors also impact predictive outcomes, as discussed in reference [3].

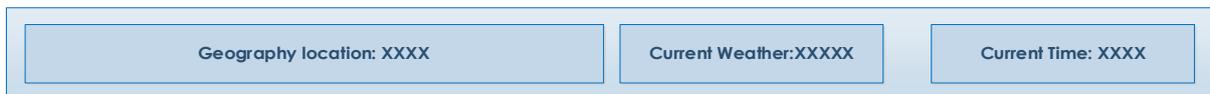

Fig.2.1 Contextual information

In this design, the lighting system is one component of the overall building assets. Figure 2.2 illustrates the icon representing lighting assets, with health status categorized into five levels, ranging from Level 1 to Level 5. These levels are visually represented by color, from red (Level 1) indicating the poorest health status, to blue (Level 5) representing optimal health. Level 5 in blue denotes the most robust condition, while Level 1 in red signifies the need for immediate attention.

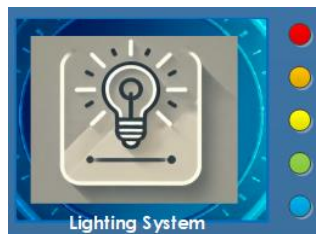

Fig.2.2 Health status of lighting assets

Similarly, Figure 2.3 presents the integrated health status of all building assets. This assessment is categorized into five levels, ranging from Level 1 to Level 5, and visually represented using color indicators. The criteria for the integrated assessment can be determined using various decision-making approaches, such as weighted averages or critical analysis of individual asset health metrics.

This integrated evaluation should be incorporated into the Building Management System (BMS) for further study and implementation. In addition to assessing the status, the design also highlights predictive results, using 3-month and 6-month projections as examples to showcase future asset conditions.

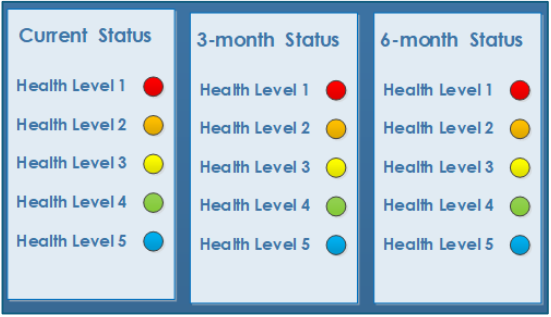

Fig.2.3 Integrated health status of all building assets

If an abnormal light is detected or if the asset manager wishes to review the status or predictive insights for any subsystem, the corresponding icon will serve as a guide, directing the manager to the relevant subsystem page for further analysis.

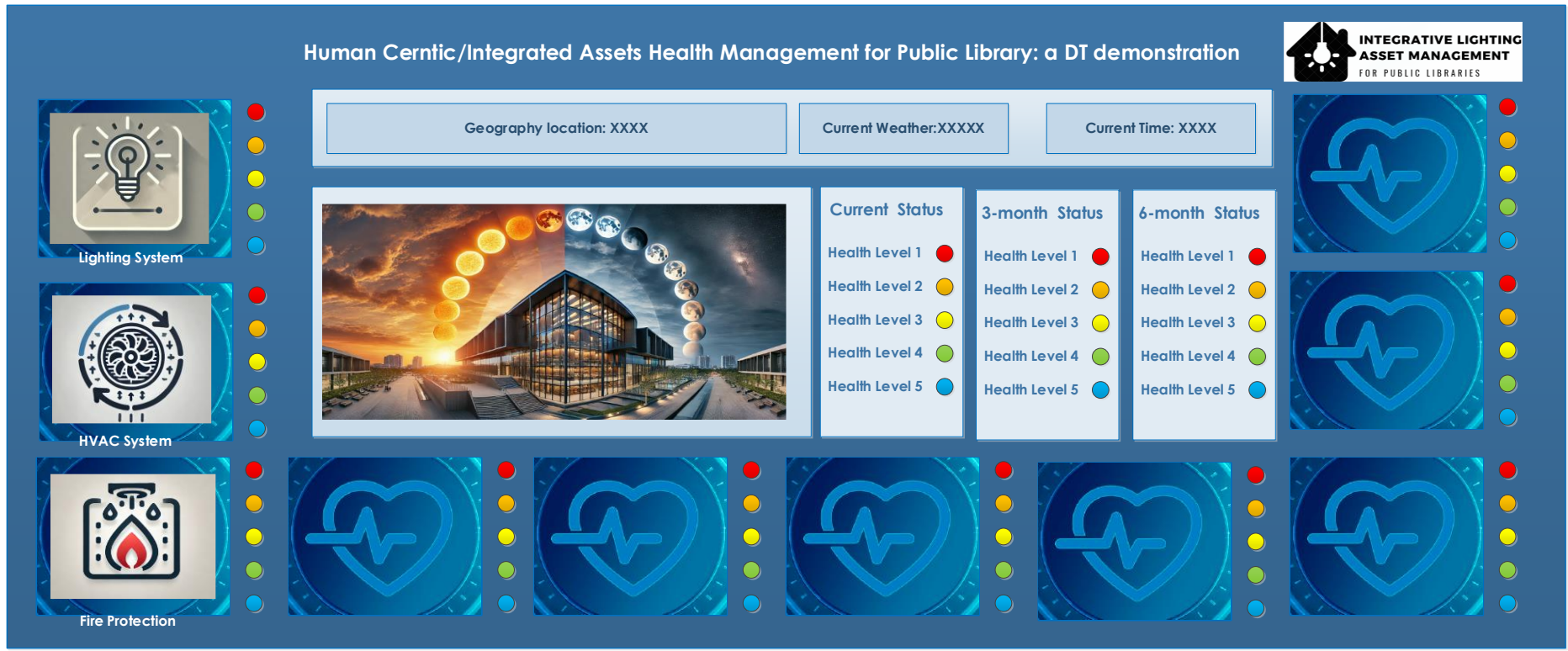

Fig.2 Health status at the system (building) level

## 3.2 Sub-system level (lighting system) Monitoring

Figure 3 illustrates the health status at Subsystem Level, using the lighting system as a focal point.

The lighting system can be displayed according to the building's structure, for instance, by floors 1 to 3 if the library consists of three levels, as shown in Figure 3.1. The overall health status is indicated by different colors corresponding to each floor.

In Figure 3.1, the left icon represents a realistic depiction of the floor, while the middle icon presents a 2D floor plan. The right icon showcases a digital demonstration in a 3D model that integrates both current natural lighting and artificial lighting considerations. Abnormal colors will be reflected in both the 2D and 3D icons in the corresponding areas.

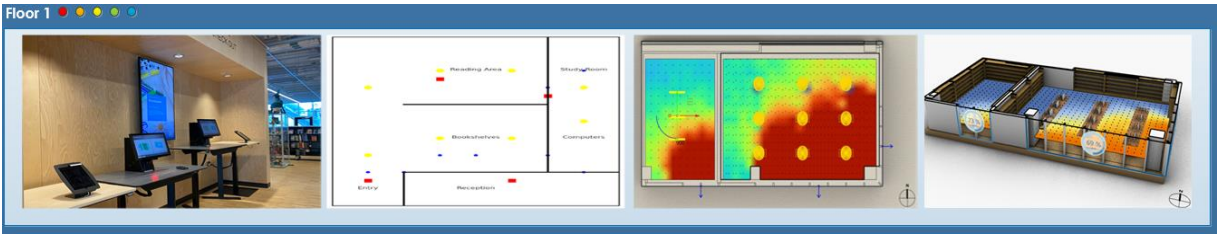

Fig.3.1 Integrated health status of lighting assets

In Figure 3.2, the icon represents various user areas on the corresponding floor. The identification of Critical Integrated Levels (CIL) is marked in the right corner of the icon. The methodology for identifying CILs is discussed in Reference [3]. The integrated status reflects the current condition, along with 3-month and 6-month predictions, each indicated by different colors.

The criteria for the integrated assessment can be established using various decision-making approaches, such as weighted averages or critical analysis of individual asset health metrics.

If an abnormal light is detected or if the asset manager wishes to review the status or predictive insights for any subsystem in a specified user area, the corresponding icon will serve as a guide, directing the manager to the relevant subsystem page for further analysis.

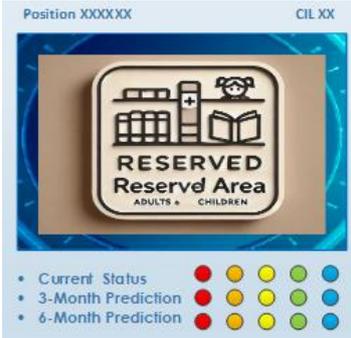

Fig.3.2 Health status of lighting assets on different user area

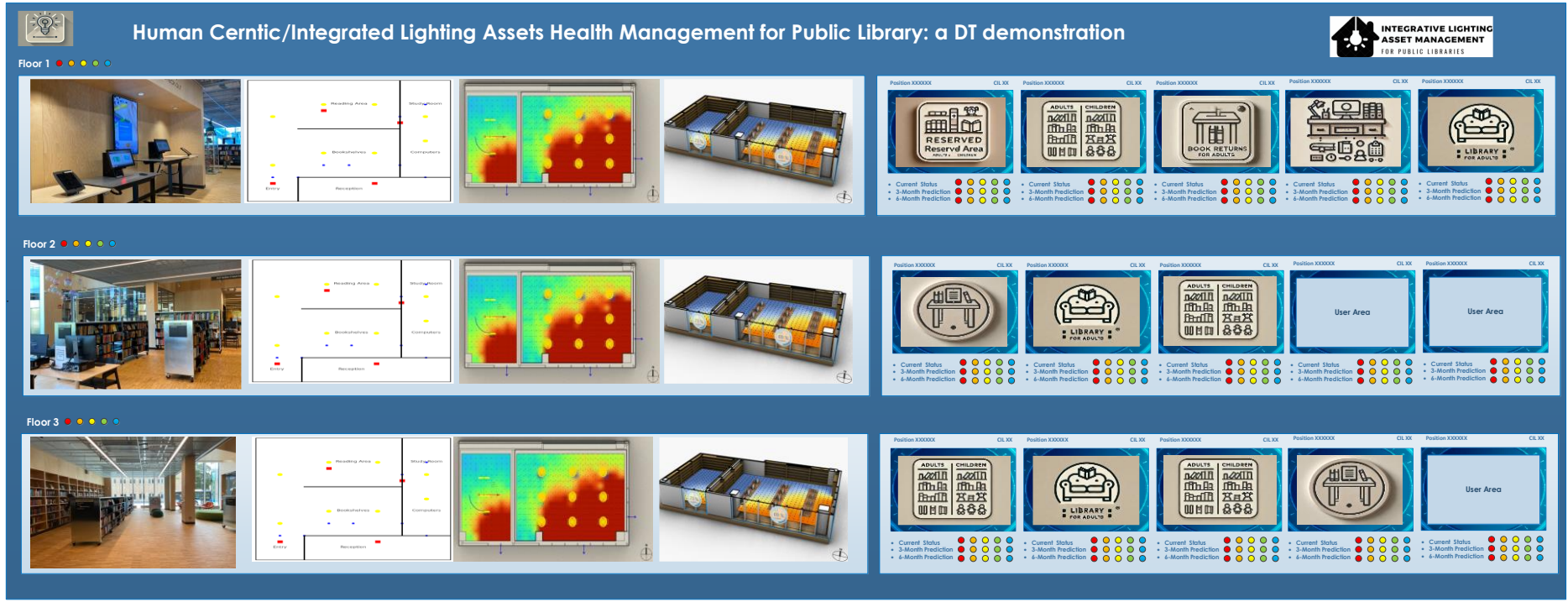

Fig.3 Health status at the subsystem (lighting) level

## 3.3 Component (user area) Monitoring

Figure 4 illustrates the health status at component Level, using the book reserve area as a focal point.

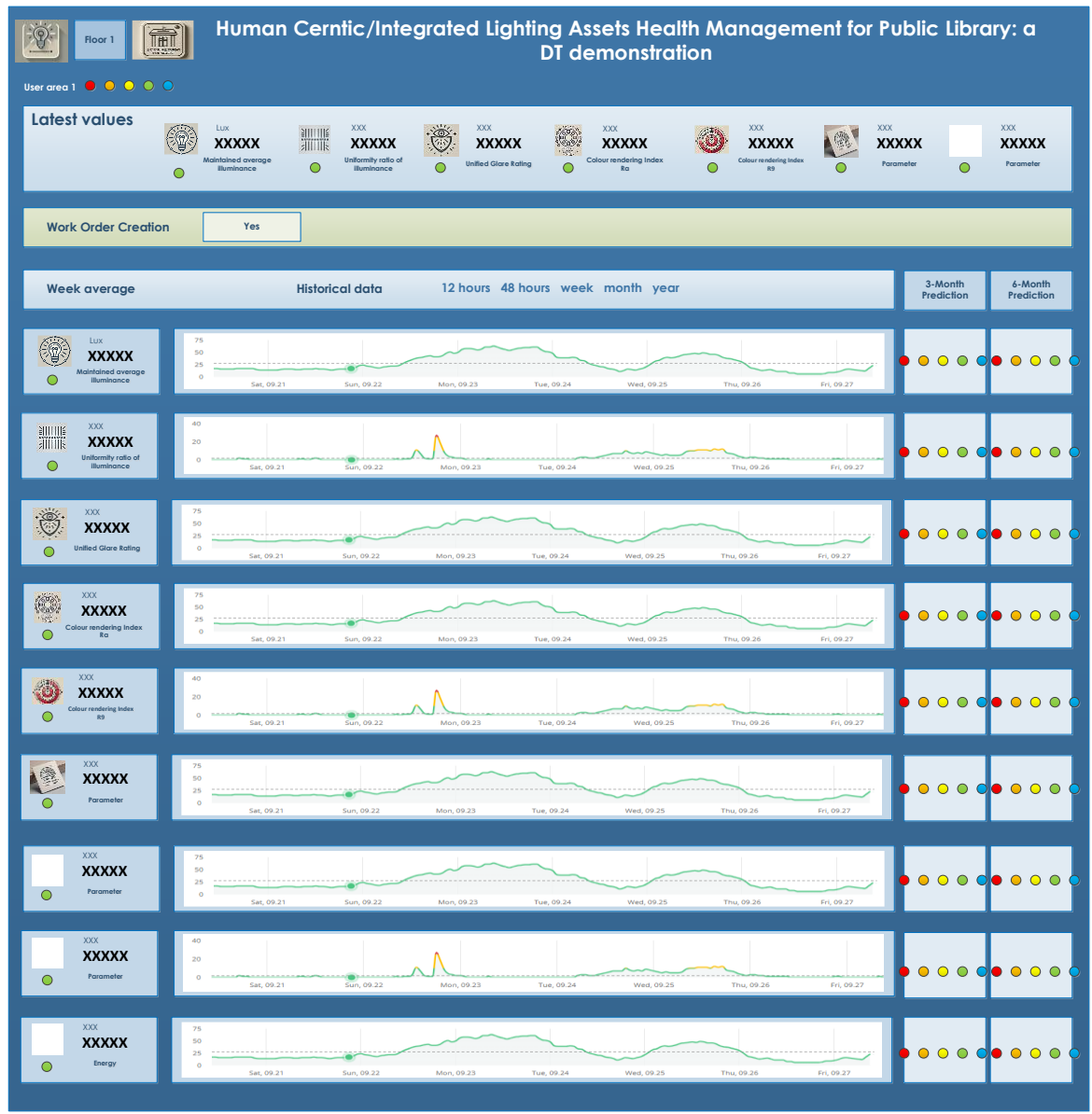

Fig.4 Health status at the component level

In this study, we identified ten parameters for assessing lighting performance across different user areas. Figure 4.1, referencing [3], displays the key lighting performance parameters along with essential information and the health status of these parameters within the focused user area.

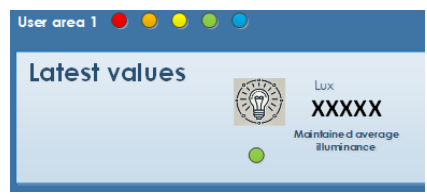

Fig 4.1 Latest values of various user area with focused lighting parameters

If any abnormalities trigger the creation of a work order at this stage, it falls under "Corrective Maintenance (CM)" work orders. In Figure 4.2, the "yes" icon indicates the initiation of the CM work order. However, the detailed CM work orders must be integrated with those of other assets, considering resource availability within the Building Management System (BMS) and applying optimization strategies for maintenance.

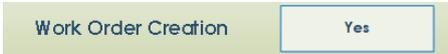

Figure 4.2 Creation of CM work orders

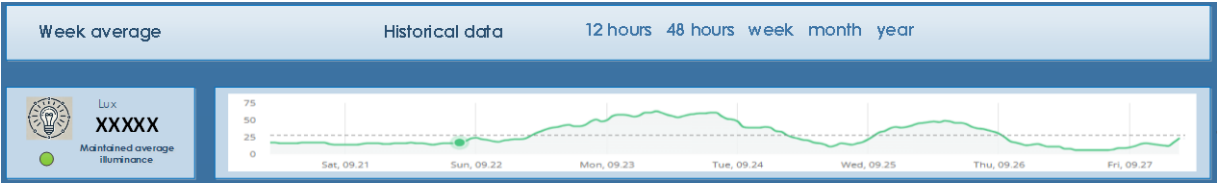

Figure 4.3 Monitoring of lighting performance parameters

Figure 4.3 displays the corresponding lighting performance parameters as defined in Reference [3]. On the left, the weekly average values are shown, with historical data customizable to view over periods of 12 hours, 48 hours, a week, a month, or even a year. The historical performance values are represented in different colors, reflecting varying health statuses from Level 1 to Level 5. The definitions of each performance status are discussed in Reference [3].

This page not only presents the values related to descriptive analytics but also provides diagnostic analytics, answering questions such as "where and why this alarm occurs." More importantly, it displays predictive values, with examples for 3-month and 6-month forecasts, see Figure 4.4.

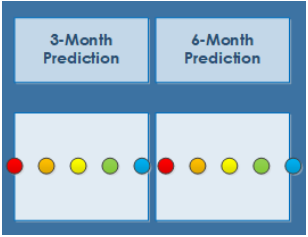

Figure 4.4 Prediction

If an abnormal light is detected, or if the asset manager wishes to review predictive insights for any lighting performance parameters, the corresponding icon will direct the manager to the relevant prediction page for further analysis. Figure 5 provides an example of a 3-month prediction, while Figure 6 illustrates a 6-month predictive analytics example.

The prediction methodologies have been discussed in detail in Reference [6].

### 3.4 Prediction and Prescription

Figure 5 and Figure 6 illustrate the prediction analytics.

If any abnormalities trigger the creation of a work order at this stage, it falls under "Preventive Maintenance (PM)" work orders or "Predictive Maintenance (PdM) work orders", see Figure 5 and 6. However, the detailed PM/PdM work orders must be integrated with those of other assets, considering resource availability within the Building Management System (BMS) and applying optimization strategies for maintenance.

The workorders reflect the prescription work in this design.

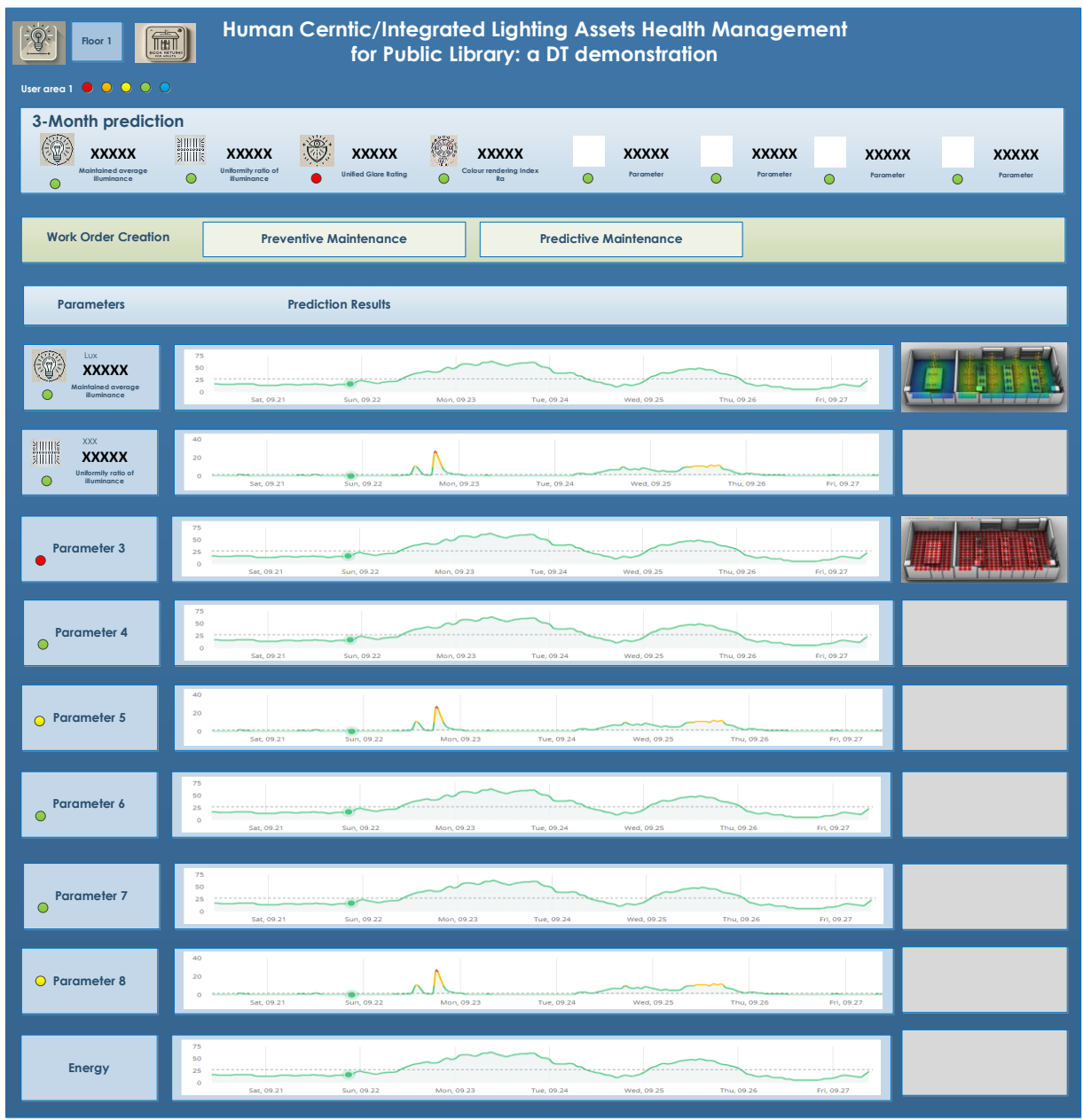

Figure 5. 3-month Prediction

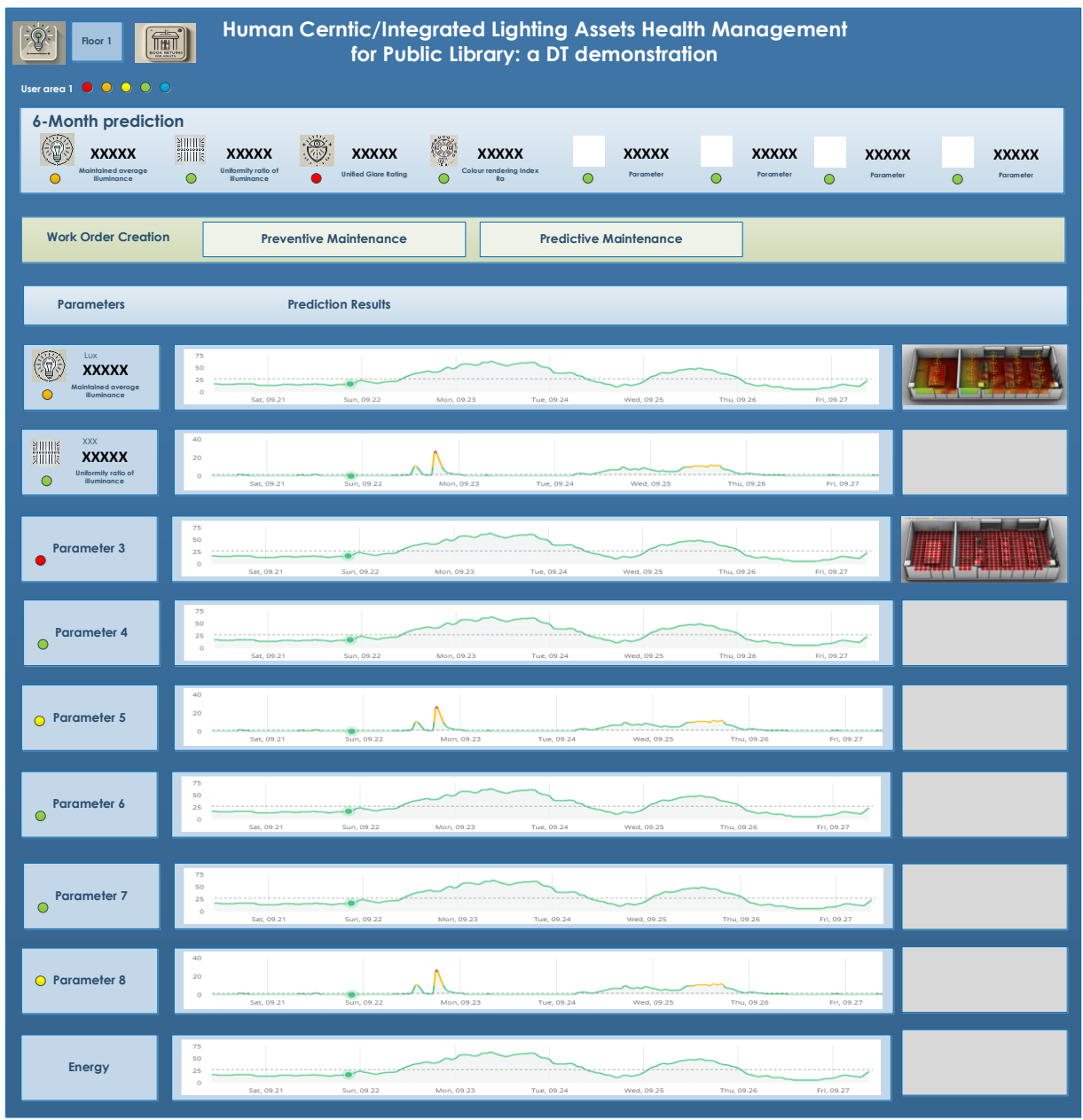

Figure 6. 6-month Prediction

## 4 Conclusions

This demonstration of an enhanced digital twin model shows a complete transition from corrective to predictive maintenance for lighting asset management in public libraries. By leveraging descriptive, diagnostic, predictive, and prescriptive analytics, the model offers a comprehensive approach to asset health management. While the design focuses on lighting systems, the framework can be extended to other building assets, offering a scalable solution for sustainable asset management in public spaces.

**Acknowledgements**

We extend our gratitude to the Swedish Energy Agency (Energimyndigheten) for their financial support, which made this research possible. The project, "Integrated Lighting Asset Management in Public Libraries (Integrerad tillgångsförvaltning för belysning i allmänna bibliotek genom Digital Tvilling)", bearing the project number P2022-00277, has benefited immensely from their backing. Special thanks are due to Dr. Jörgen Sjödin, the project manager at the Swedish Energy Agency, whose expert guidance has been indispensable throughout the research process.